%% file: schmidt-fest.tex
\documentclass[submission]{eptcs}
\providecommand{\event}{Schmidt Festschrift 2013} 

\usepackage{verbatim}
\usepackage{graphicx}
\usepackage{wrapfig}
\usepackage{url}

\title{The Immune System: the ultimate fractionated cyber-physical system}
\author{Carolyn Talcott
}
\def\titlerunning{IS as FCPS}
\def\authorrunning{C. Talcott}
\begin{document}
\maketitle

\begin{abstract}

In this little vision paper we
analyze the human immune system from a computer science point of view
with the aim of understanding the architecture and features that 
allow robust, effective behavior to emerge from local sensing and actions.
We then recall the notion of fractionated cyber-physical systems, and
compare and contrast this to the immune system.  We conclude with
some challenges.

\end{abstract}

\input prelude


\input introduction


\input is

\input design


\input fcps

\input is-v-fcps



\input conclusion


\bibliographystyle{eptcs}
\bibliography{local}
\end{document}

%% file: prelude.tex
\section*{Prelude}\label{prelude}

It is an honor and a pleasure to be part of the Festschrift for Dave. I have
known Dave for quite a long time, probably going back to the Atlantique
US-European project that I co-organized with Neil Jones.
Dave brings a special humor to discussions (and celebrations) along with deep
and key insights. Over the years we have shared interests in programming issues:
languages, semantics, transformation and specialization, analysis \ldots. I
always look forward to chatting when our paths cross, which fortunately is reasonably
often.

This is my second opportunity to give a talk honoring Dave. The first was in 1998 when
I gave a talk on actors at a luncheon honoring Dave at KSU. Going beyond actors and
continuing in the spirit of tackling messy problems, and encouraging my colleagues to
think about the emerging challenges, here I consider the notion of fractionated
cyber-physical systems, analyzing the human immune system to see what can we learn
from this amazing system for design and understanding of a next generation of
cyber-physical systems.  I envision that study of the immune systems as
a cyber-physical system will give rise to new programming abstractions and
interesting challenges for program abstraction needed to analyze actual systems.

I hope that the reader, and Dave in particular, is entertained by the analysis
of the immune system as well as finding food for thought, challenges and maybe
new insights.

%% file: introduction.tex
\section{Introduction}\label{introduction}

Computer enabled systems are becoming both ubiquitous and increasingly complex,
moving from isolated embedded control systems to open interactive systems with
essential integration of the cyber and the physical, hence the term
``cyber-physical'' system. Such systems are formed from distributed components
of diverse capabilities that interact with an unpredictable environment. One
example is modern cars. They are not only concerned with controlling operation
of engine, brakes, locks, and such, but also with helping the driver be aware
of surroundings (other vehicles, people, obstacles), entertainment, and with
monitoring system function and tracking-maintenance. Other examples include
unmanned vehicles, smart buildings, assisted living, medical devices,
manufacturing (including 3-d printing), and emergency response assistance. In
addition, there is an explosion of small apps running on mobile devices. We can
envision a future in which these apps combine and collaborate to provide
powerful new functionality not only in the realm of entertainment and social
networking, but also harnessing the underlying communication and people power
for new kinds of cyber crowd sourcing tasks.

How do we design, build, and understand such systems? The actor model
\cite{baker-hewitt-77cpp,agha-thesis,agha-mason-smith-talcott-96jfp} was an
important step in computational models for open distributed systems. The key
ideas included independent computational agents, with secure reliable
point-to-point communication, and a causal ordering on events based on
physics---something can only be received after is sent. The actor
communication model is both a strength and weakness. It provides a level of
abstraction that enables formal analysis of system behavior. However this level
of abstraction hides details that are important for designing systems that must
be resource aware and function in unreliable, unpredictable environments,
and in which space and time matter.  New models of computation and interaction
are needed.

The human immune system \cite{works-3,janeway-7} is a fascinating example of a
complex, robust, adaptive system. It has been studied from an 
artificial intelligence perspective, and
from the intrusion detection network security perspective. In this paper we
analyze the immune system from a broad computer science perspective to identify
building blocks, mechanisms, and features that are key to the successful
operation and indicate analogs to diverse computer science concepts (section
\ref{is}). We abstract from the analysis to outline an architecture and design
principles for immune-like CPS (section \ref{design}).  We then recall the
notion of fractionated cyberphysical system introduced in
\cite{stehr-etal-11cltfest} (section \ref{fcps}), and compare and contrast 
the NCPS framework with the Immune System  (section \ref{is-v-fcps}). 
We conclude with some future directions and challenges 
to enable deeper understanding of these systems (section \ref{conclusion}).

%% file: is.tex
\section{The Immune System as a CPS}\label{is}

The human immune system consists of billions of cells distributed throughout
the host body. Each cell operates autonomously based on local information
sensed from its environment and communication with its neighbors. The job is to
identify, contain and eliminate invaders (aka pathogens) without damaging the
host. This requires the coordination and controlled action of cells with
different capabilities. The fact that normally, the desired behavior emerges is
completely amazing. We expect that there are new design principles and notions
of control theory that can be inferred by studying this system from a
computational point of view. The following is base on two classic immunology
texts: a gentle introduction \cite{works-3}, and a standard graduate immunology
text \cite{janeway-7}.

\begin{figure}[ht]
\centering
\includegraphics[width=0.8\linewidth]{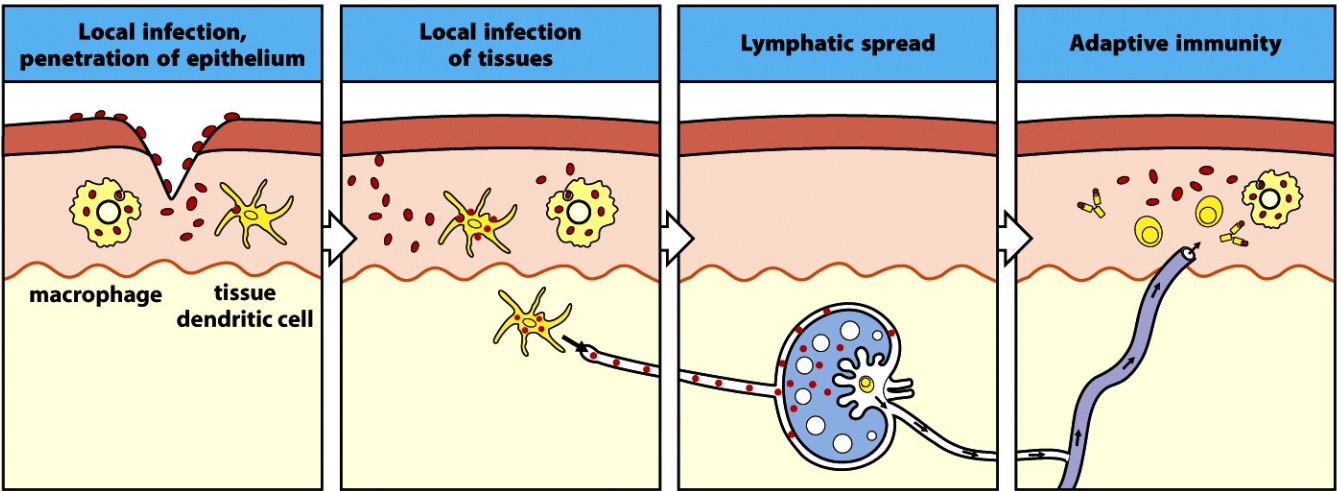}
\caption{\label{scenario}
The immune system in action}
\end{figure}

We begin with a little scenario (Figure \ref{scenario}
from \cite{janeway-7} Chapter 10 Figure 2) showing the immune system in
action. 

The players in act I (the leftmost two panels of Figure \ref{scenario}) are
\emph{macrophages} and \emph{dendritic} cells, and of course the pathogens (red
dots). These cells are members of the \emph{Innate Immune System}, the first
line of defense. Macrophages (Greek: big eaters, from makros ``large'' +
phagein ``eat'') are generally in charge of garbage collection/recycling. When
they encounter a pathogen their job is two fold, to engulf and eventually
destroy the pathogen and to send signals indicating the presence of pathogens
in order to alert near by cells and to attract help. Dendritic cells get their
name from their from their tree like branched shape (dendron is greek for
tree). A dendritic cell continually samples its surroundings. When it
recognizes a pathogen, information about the pathogen and other environment
factors is summarized and presented on its surface and the cell travels through
the lymph system to carry the word (second panel of Figure \ref{scenario}).

There are many more types and subtypes of immune system cells. A key player in
the innate immune system not shown in the scenario is the \emph{neutrophil}.
Neutrophils patrol in the blood stream and migrate to infection sites. Their
job is clearance of extracellular pathogens (cytotoxic T cells, see below, take
care of intracellular pathogens). A neutrophil is activated in response to
detecting common components on the surface of pathogens together with a signal
for help. Once active, a neutrophil migrates quickly to the site of infection
(within minutes as opposed to response time of days for the adaptive immune
system). A neutrophil can engulf pathogens tagged by antibodies, bringing them
into a microbicidal environment. It can also emit granules that that dissolve
and release antimicrobial toxins. Finally, a neutrophil can extrude webs that
trap and kill microbes. Another important player in the innate immune system is
the natural killer cell (NK cell). An NK cell becomes active when it detects
presence of pathogens. Its job is to kill damaged/defective cells (for example,
cells harboring replicating pathogens, or behaving aberrantly) either by
injecting toxins or by sending a signal that causes the target cell to commit
suicide. It uses a two signal process to decide whether or not to kill. A kill
signal is activated by detecting abnormal patterns on the target cells surface,
indicators of some damage. A don't-kill signal is activated by detecting a
healthy status display on the target cell (MHC I, see below). Clearly a
potentially dangerous weapon.

\begin{figure} [hg]
\hfill
\begin{minipage}{2.1in}
\includegraphics[width=\linewidth]{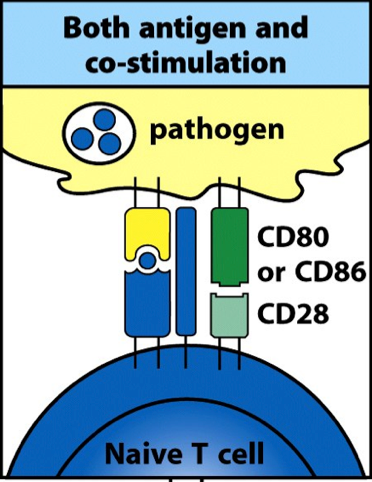}
\caption{\label{tcell-act} T cell activation}
\end{minipage}
\hfill
\begin{minipage}{2.0in}
\includegraphics[width=\linewidth]{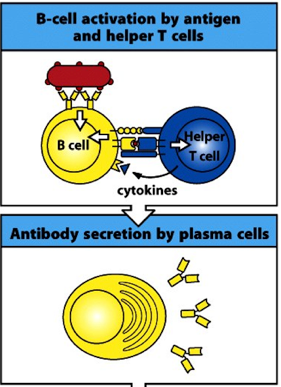}
\caption{\label{bcell-act} B cell activation}
\end{minipage}
\hfill
\end{figure}

In act II (the two rightmost panels of Figure \ref{scenario}) members of the
\emph{Adaptive Immune System}, including \emph{T cells} and \emph{B cells},
join the battle. Instead of recognizing general classes of pathogen, T and B
cells recognize signatures of specific pathogens. In particular, for each
possible pathogen signature there are T and B cells around that recognize this
signature. However they don't respond without additional authorization. A
dendritic cell presenting a pathogen signature meets with T cells in a lymph
node. As illustrated in Figure \ref{tcell-act}, a T cell with receptor that
recognizes the signature pattern presented by the dendritic cell will initiate
activity if it receives an additional authorizing signal from the dendritic
cell (via a co-receptor). In particular, the T cell receptor (blue plug)
matches the pathogen signature (yellow plug), and is authenticated by matching
the CD28 co-receptor pattern (light blue plug) to a surface CD80 or CD86 (green
plug).


As shown in Figure \ref{bcell-act}, a B cell first recognizes the whole
pathogen (red shape binding to the yellow Y shaped receptor), degrades it and
presents the resulting signature in a manner similar to the dendritic cell
(yellow plug pointing to the right). It is not yet fully active. One way to
become active is to meet an active T cell that recognizes the presented
pathogen signature. After authenticating (matching of yellow beaded co-receptor
to the blue pattern), the T cell will send a signal to activate the B cell. Now
the active T and B cells can go to work. Active T and B cells send further
specific signals to communicate status to other cells. They also create copies
of themselves, specialized to recognize the same signature, revving up the
defense. B cells mainly circulate in the blood stream (plasma B cells) and
generate \emph{antibodies} (the Y shaped icons in the lower panel of Figure
\ref{bcell-act}). Antibodies generated by an active B cell recognize the the
same full pathogen pattern as the parent cell. They circulate and attach
themselves to pathogens that they recognize, thus preventing pathogenic effects
and making the pathogen attractive to destructors.

There are two main types of T cells: called \emph{helper} and \emph{effector}.
Helper T cells, in addition to activating B cells, emit signals that control
the activity of other types of cells. The job of effector T cells, also known
as cytotoxic T cells (CTCs), is to kill cells that have been damaged somehow,
for example cells that have been infected by pathogens that are replicating
inside. (Cancer cells can also be attacked by CTCs).

While infection remains, the T and B cells are stimulated to continue activity.
When the infection is cleared the active cells loose activity and die.
Some activated B and T cells will stay on the sidelines, becoming memory cells. 
These cells can be reactivated quickly if there is a later infection.
(This is basically how vaccination works and why you are unlikely to get
measles or mumps twice.)    

\emph{How are pathogen signatures presented?} Some immune cells want to know
about pathogens in the cellular environment while others want to know about
pathogens living and replicating inside a host cell. How can this be done? The
answer is MHC (major histocompatibility complex) classes. MHCs are protein
complexes used by cells to display protein fragments (such as pathogen
signatures) on their surface. MHC class I (MHC-I) presents fragments resulting
from digesting material inside the cell. CTLs, looking for internal infection
scan MHC-Is. MHC class II (MHC-II) presents protein fragments digested from the
environment. Presentation by MCH-IIs initiate activation of T cells, and helper
T Cells look for MHC-II presentations by B cells to complete B cell activation.

\emph{How do CTCs do their job?} An active CTC must migrate to the site of
damaged cells and determine which cells are damaged. Every cell makes a summary
of its internal state (a selection of digested peptides) which is carried to
the surface by the MHC-I mechanism discussed above. If the cell is infected,
the degraded pathogen's signature will be presented on the surface of the cell.
A CTC specialized for this pathogen will recognize the infected cell. It can
kill the cell either by perforating the cell surface and injecting proteins
that will generate a suicide signal (apoptosis). The CTC can also send the
suicide signal directly via a surface receptor on the target cell that
recognizes a protein on the surface of the CTC. The use of MHC-I presentation
focuses CTC capability (expensive) on infected cells (groups of viruses) while
antibodies (which are plentiful and cheap) can take care of single free
viruses.

\emph{How do T and B cells gain specificity?}  
There is a unique T or B cell receptor and antibody type for each organic
compound. Each mature T and B Cell (and its progeny) produces exactly one type
and (almost) every type is produced by some T or B cell in the system. How can
this be? There are not enough genes. The trick is that the DNA of immature B
and T cells is specialized in the process of maturation. Initially the DNA has
multiple instances of several modules. These modules combined in a series of
clip/rejoin operations resulting in one of the many possible combinations. In
general each maturing cell picks a new combination.

\begin{figure}[ht]
\centering
\includegraphics[width=0.67\linewidth]{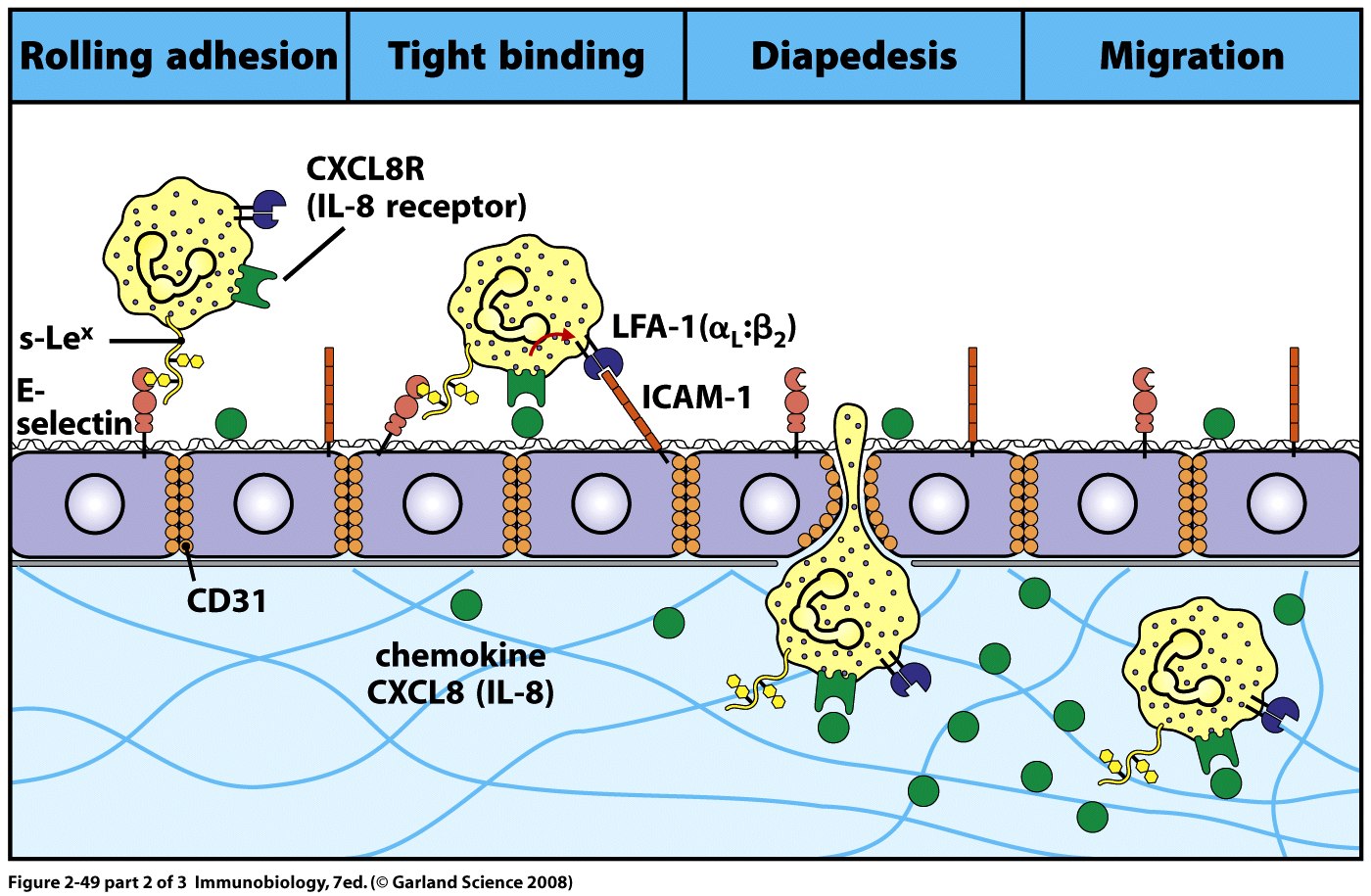}
\caption{\label{exit} Exit (from \cite{janeway-7} Chapter 2 Figure 49)}
\end{figure}  

\emph{How do traveling immune system cells know where to go?} Cells at an
infection site emit signals (cytokines and chemokines) that tell the blood
vessels to setup an exit point. The vessels become more permeable and they
expose proteins that act as signs and hooks. A cell looking for an infection
site exposes proteins that recognize the signs and bind to the hooks, bringing
them to a halt. Then they slip through the vessel wall, attracted by the
infection signals. (See figure \ref{exit})

\emph{What can go wrong?}
Many things can go wrong. Mostly resulting in collateral damage to
the host system (referred to as self).  Here are two classes.
Among the many combinations of modules to form T or B cell receptors are those
that recognize patterns on the surface of the host/self cells. Activating such
cells would result in attack of self. Not good. In addition, some combinations
result in non-functional proteins (they can not recognize their target
patterns), leading to useless cells. Thus maturing T cells undergo
\emph{tolerance} tests to ensure they do not recognize self and
\emph{competence} tests to ensure that the resulting protein is functional.

There are several types of cells in the innate immune system whose job is to
kill damaged cells or pathogens. This often involves emitting non-discriminant
toxins. Neutrophils are an example. If they get out of control they can cause
substantial damage to self. The short life of neutrophils is one form of
control. An active neutrophil can only do limited damage in its short life, if
it fails to find pathogens to attack. If neutrophil eats a pathogen and fails
to kill it, the neutrophil becomes an incubator and the short life limits the
time that pathogens can grow and replicate inside.

\emph{Importance of signaling!} Signaling is a crucial aspect of how the immune
system cells coordinate and achieve a balance that allows clearance of
intruders with minimal collateral damage. Cells communicate either by
presenting information on their surface (such as the MHC presentation
mechanism, or co-receptors and matching ligands needed for activation of T and
B cells) or by secreting cytokines (Greek cyto-, cell; and -kinos, movement)
and chemokines (small chemotactic cytokines with the ability to induce directed
chemotaxis in nearby responsive cells). Each cytokine has a matching
cell-surface receptor. Signal reception leads to cascades of intracellular
signaling that may alter a cell's behavior, for example, causing more/different
cytokines to be secreted, receptors for different signals to be exposed,
repression of production of signals and receptors, increase/decreased rate of
proliferation, or even cell death.

One thing that makes the signaling process tricky is the feedback loops. When a
natural killer (NK) cell sees Lps, a conserved pathogen surface molecule, it
becomes active and secretes Ifng (interferon gamma). When a macrophage that has
also seen Lps receives the Ifng signal it secretes the Tnfa (tumor necrosis
factor alpha) signal. When the macrophage sees the Tnfa signal it secretes IL12
(interleukin 12). When a NK cell receives both Tnfa and IL12 it responds by
secreting more Ifng and also secreting IL2 (interleukin 2), a signal to
proliferate. It can also now receive the IL2 and starts proliferating, making
more NK cells to produce Ifng and IL2! Another example is the signaling between
macrophages (Mph) and helper T cells (T0) which can choose to be Th1 or Th2
cells, each with distinct signaling profiles. 

We conclude this section with hints as to how formal models that capture
the biologists intuitions and diagrams can help to understand how the immune
system works. First, a little scenario \'a la
Bob and Alice security protocols to illustrate the interactions, including
positive and negative feedback. A $\rightarrow$ B: C says that A sends a
message C (for example a cytokine) that is intended for (can be received by) B.

\begin{itemize}
\item
  Mph $\rightarrow$ T0: IL12  (become Th1)
\item 
  Th1 $\rightarrow$ Mph: Ifng (keep signaling)
\item 
  Th1 $\rightarrow$ Th1: IL2  (proliferate)
\item 
  Th2 $\rightarrow$ Th2: IL4  (proliferate)
\item 
  Th1 $\rightarrow$ Th2: Ifng (don't proliferate)
\item 
  Th2 $\rightarrow$ Th1: IL10 (don't proliferate)
\item 
  Th2 $\rightarrow$ Mph: IL4  (don't make IL12)    
\end{itemize}
In summary, we have
\begin{itemize}
\item
  IL2 causes proliferation, up-regulation of the IL2 receptor and production of 
  more IL2
\item
  Ifng causes production of Tnfa and IL12 which in turn causes more Ifng
\item
  Ifng suppresses production of IL10 and IL4 (by suppressing Th2 activity)
\item
  IL10 suppresses Ifng (by suppressing Th1 activity)
\item
  IL4 decreases IL12 production by Macrophage, leading to fewer T0 cells
  becoming T1 cells.
\end{itemize}

Activation/inhibition relations such the above scenario can be
modeled using boolean networks and analyzed for possible steady
state properties (qualitative features such as attractors, positive/negative
feedback loops).  

Next, sample rules from a
rewriting logic model of the immune system implemented in the Maude language
\cite{clavel-etal-07maudebook}.  These rules contain
more detail than the Alice and Bob style interaction
sequence relations, but still omit many low level details.   The system state
is represented as a multiset of cell objects of the form

{\small
\begin{verbatim}
        {loc | celltype - pmods smods xmods}
\end{verbatim}
}

\noindent The symbol \texttt{loc} stands for the location, for example
peripheral tissue (\texttt{PTS}) or lymph node (\texttt{LN}).
The symbol \texttt{celltype} names the type of cell: 
macrophage (\texttt{Mph}),
dendritic cell (\texttt{DC}),
Helper T Cell (\texttt{TC4}, \texttt{TH1}),
or pathogen \texttt{Path}.  The three modification attributes
\texttt{pmods}, \texttt{smods}, and \texttt{xmods} represent
different aspects of a cells state.
\texttt{pmods} are phenotypic description such as
\texttt{resting}, \texttt{naive}, \texttt{active}, \ldots,
\texttt{smods} are proteins being secreted,
prefixing the protein name with an \texttt{s}.
\texttt{xmods} are proteins expressed on
the cells surface, prefixing the protein name with an \texttt{x},
including receptors, co-receptors, and various probes/hooks.   
A rule has the form 

{\small
\begin{verbatim}
label:
  state-before
  =>
  state-after .
\end{verbatim}
}

\noindent
In the rules a cell state also includes a variable such as \texttt{macmods} or
\texttt{dcmods} to match
any modifiers of a concrete cell object that are not relevant
for the rule.

The rule \texttt{014} shows a resting macrophage 
\texttt{[Mac - macmods resting]} 
in the peripheral tissue (\texttt{PTS}) encountering  Lps coated
pathogens \texttt{Path}.  The macrophage engulfs the pathogen and presents
the result (\texttt{xMhcI*,xMhcII*}). It also now secretes
Tnfa (\texttt{sTnf})

{\small
\begin{verbatim}
rl[014.Mac.exposed.to.Path]:
  {PTS | pts Path [Mac - macmods resting]                            }
  =>
  {PTS | pts Path [Mac - macmods presenting sTnf xMhcI* xMhcII* xB7] } .
\end{verbatim}
}

\noindent
A dendritic cell also observes the pathogens, digests the pathogen and travels to a
lymph node (rules omitted). Here it meets a naive T cell (\texttt{TC4}) that
recognizes the presented pathogen signature (rule 008). The DC secretes IL12 causing
the T cell to differentiate into a helper T cell of type 1 (TH1). The TH1 cell up
regulates its IL2 receptor \texttt{xIL2Ra.hi} and also secretes IL2 and
Ifng. The IL2 will induce replication. 

{\small
\begin{verbatim}
rl[008.TC4.becomes.TH1]:
  {LN  | ln  ([TC4 - tc4mods naive xIL2Ra.lo] : 
              [DC - dcmods mature xMhcII* xB7])  }
  =>
  {LN  | ln  ([TH1 - tc4mods primed sIL2 sIfng xIL2Ra.hi xVLA4 xFas xFasL] :
              [DC - dcmods mature xMhcII* xB7 sIL12 ])                      } .
\end{verbatim}
}

\noindent
The TH1 cell further matures to become \texttt{effective} and travels to
the infection site (rules omitted).   Here the TH1 cell meets 
a macrophage that has engulfed the same pathogen and is \texttt{presenting}.
The two cells recognize each other via the Cd40-Cd40L receptor
ligand match.

{\small
\begin{verbatim}
rl[018.TH1.Mac.effects]:
  {PTS | pts ([TH1 - th1mods effective] :
              [Mac - macmods presenting xMhcII*]) }
  =>
  {PTS | pts ([TH1 - th1mods effective xCd40L sIfng] : 
              [Mac - macmods active xMhcII* xCd40 xTnfRs])           } .
\end{verbatim}
}

\noindent
The Ifng secreted by TH1 cell increases  the antimicrobial effectiveness 
of the macrophage, enabling it to kill the pathogen inside.
The TH1 cell stops  secreting Ifng, and the macrophage returns
to a resting state.

{\small
\begin{verbatim}
rl[019.Mac.act.by.TH1]:
  {PTS | pts ([TH1 - th1mods effective xCd40L sIfng] : 
              [Mac - macmods active sTnf xMhcI* xMhcII* xCd40 xTnfRs]) }
  {Sig | sig                                                           }
  =>
  {PTS | pts  [TH1 - th1mods effective] [Mac - macmods resting]        }
  {Sig | sig  INTERNAL-PATH-DEAD                                       } .
\end{verbatim}
}

Models such as the rewriting logic model of the immune systems provide a way of
organizing the available information at a meaningful level of detail. They can be
executed, and the execution space can be searched for different possible scenarios.
Model-checking can be used to find executions that lead to states/phenotypes of
interest \cite{talcott-08sfmbio,lincoln-talcott-10ssb}. Rewriting models serve
as a starting point for understanding how the control mechanisms work.
They can be abstracted to boolean networks for abstract dynamical systems analysis.
Quantitative information such
as rates, concentrations, or probabilities of interactions can be added to allow
simulations and different forms of statistical or probabilistic analysis.

%% file: design.tex
\section{Features and Principles of the Immune System}\label{design}

The job of the immune system is to disable and/or destroy invaders/pathogens. As
indicated in section \ref{is}, it is a very complex system, which generally
works quite well. What are the building blocks and design principles that could
explain this? Here we make a small start at answering that question.

\subsection{Features}
A two level architecture organizes the players according to specificity of
detection and cost/risk of action. The Innate Immune System (IIS) level consists of the
players with a general notion of target. It includes border guards (such as
macrophages and dendritic cells found near the surfaces) and troops on patrol
(such as neutrophils in the blood stream). These players provide early defense
and alert the other players, seeking out those with the right specificity. The
Adaptive Immune System (AIS) level consists of the players with highly specific
recognition and actions. They can be highly aggressive and safety mechanisms are
needed to prevent collateral damage. Control signals (battle alert system) are used to
initiate, stimulate, and turn off activity.

On the cyber side, all the cells have the same ``program'' (DNA). They evolve to
play distinct roles (called differentiation) according to location and
environmental cues. Some specialize their behavior further by choosing a
particular pattern (pathogen signature) to respond to. This leads to
a diversity of possible behaviors and being prepared for all possible pathogen 
signatures.

On the physical side location is important. Cells can only communicate when they
are co-located, and they must move to an infection site to act on pathogens.
Information propagates locally by signaling molecules (such as cytokines and
chemokines) similar to wireless broadcast. Long range propagation relies on
mobility of couriers.  Cells are also resource limited.  Creation of new cells
uses resources generated by metabolism and recycling and requires energy sources.

\paragraph{Self evaluation, monitoring and maintenance.}
All cells provide a summary of their internal state using MHC-I presentation.
Other cells, for example CTC cells, can scan the summary to determine the cells
health status. The immune system vets maturing T Cells to make sure they are
safe and functional.

All cells have a finite lifetime, dead cells are recycled, and new cells are
continuously being produced. Cells die naturally, commit suicide, or are killed
for any of several reasons: they are no longer needed, timing out stale
information, or they are worn out, damaged or defective. New cells are produced
in fresh undifferentiated form to ensure a continuing supply of diverse
behaviors. New cells are also produced by replication to increase the supply of
a specific behavior (active T or B cells).

\paragraph{Pattern matching and binding is a key mechanism of interaction and
action.} Patterns are built from the basic building blocks of proteins, amino
acids, and modifications thereof by complex combinations of sugar moieties.
Matching can be very specific and may depend
not only on the molecular formula, but also the 3 dimensional arrangement
in space (folding). There are patterns that characterize broad classes of
pathogens (conserved patterns), and patterns that correspond to very specific
pathogen strains (we have called these signatures). These patterns are used to
recognize a pathogen in order to take further action---eating, coating with
antibodies, pouring toxins on or into them.  
Receptors use pattern matching to recognize ligands and initiate signaling
processes. 
Other patterns are used for navigation. Cells circulate, patrolling or moving
from birth place to job site. Pattern matching identifies suitable exit
locations depending on the cell type and state (which is characterized by
patterns on its surface) and patterns exposed by vessel walls.

\paragraph{Pattern matching is the foundation for security and safety
mechanisms.} The receptor patterns a cell exposes on its surface determine the
messages/signals it can receive. What a cell exposes depends on its type and
state. Signals received are generally translated into internal signals that
eventually lead to a choice of actions/behaviors. This very specific pattern
matching plays a role similar to crypto primitives in providing aspects of
security, including authentication, and access control/need to know. 

Although the lack of cryptography may make the security model seem weak, within
the context that it is intended to work, it is quite strong. An attacker would
need to forge patterns involved in bio-security mechanisms. These patterns are
inherent in the nature of the entities involved and serve a role similar to
finger or voice prints. It works because forgery is unlikely, although
occasionally achieved for some patterns by rapidly adapting mutating entities
such as viruses. Of course biologists can and do forge such patterns as they
carry out experiments to study cellular function and design microbicides. But
that is not part of the attack model of the immune system.

Combined with the MHC mechanism pattern matching also provides provenance. The
presenter and receiver must authenticate---by additional matching of MHC type
to receiver surface proteins and possibly confirming the MHC is native to the
individual host (a source of transplant rejection). MHC-I presentation
guarantees the presented peptide comes from inside cell while MHC-II
presentation guarantees it was found in the surroundings. Of course some cells
may present both cases (an infected macrophage for example).

Competence and tolerance tests reject patterns that would lead to undesirable
behavior. The T cell activation mechanism involves three pattern matches: The T
cell must recognize the presented signature pattern, the presenter, and the
presentation mechanism. B cells activation has 2 phases. The B cell must first
recognize its pathogen, then it must connect with a T cell that has seen the
same pathogen that is exposing the necessary authentication pattern.

\subsection{Principles}

\begin{description}
\item[Diversity.]
There should be sufficient diversity to meet all needs. Maturing T and B cells
generate receptor combinations to match any possible pathogen signature.
The should be multiply means to achieve a goal. the innate and adaptive
immune systems provide alternate approaches to clearing infections.
Internal signals can generally propagate by multiple pathways.

\item[Principle of clonal selection.] 
Only cells that are known to combat a current threat proliferate.


\item[Specificity of recognition and action.] There should be both general and
specific pattern recognizers and actions. Conserved patterns that are essential
to survival of pathogen make it difficult for a pathogen to escape detection.
Patterns specific to pathogen strain allow more effective countermeasures.
Acting on conserved elements makes it difficult for pathogens to escape
destruction. However this may also damage the host. Specific targeted action is
more expensive, subject to delay, but more effective and with less collateral
damage.
\item[Safeguards.]
\emph{Multi-ary control} is characteristic of immune system decisions: entities
often must combine different signals from different sources before acting. One
example is neutrophils that need a signal reporting damage and a signal
reporting invasion before becoming active. This means that they don't respond to
the tissue damage of antiseptic surgery, or to the non-damaging bacteria that
live in the gut. The two stage activation of B cells is another example. There
should be \emph{confidence} in the meaning of a signal, before acting. An
example is the requirement of a co-receptor signal for activating a T cell. MHC
displays allows different cells to act on detection of a pathogen signature
depending on the provenance. Naive T and B cells have high activation
thresholds, there must be a sufficient threat to take action. Activity can be
sustained and/or resumed at lower signal strength.

\item[Fault tolerance.]  
Faults are expected. There is continual checking and fitness tests as
immune system cells mature, in order to detect potential harmful behavior
(for example, programmed to attack self, or weak pattern recognition).
Surviving rogue cells are prevented from action by the improbability
of seeing the right combination of signals.
Cells under control of an attacker (replicating pathogen inside) are
detected and destroyed by patrolling killer cells.

\item[Mobility.] Because location matters, some cells travel to carry a signal,
some cells travel on patrol, some travel to sites where there actions are
needed. Navigation uses potential functions and patterns to match need to
capability.
\end{description}

%% file: fcps.tex
\section{Fractionated Cyber-physical Systems}\label{fcps}


A \emph{fractionated cyber-physical system} (FCPS) \cite{stehr-etal-11cltfest}
is characterized by many small individual cyber-physical entities (CPEs)
cooperating and/or competing to achieve some goal. Imagine replacing a huge
satellite by many small independent units (cube-sats). They are inexpensive to
deploy, it is easy to add new functionality or replace a broken one. The
collection of units is agile and can form complex functional units as needed.
Other examples include micro manufacturing, systems of smart buoys, emergency
response entities. Adding humans in the mix we get new kinds of crowd sourcing
and pervasive games.

A CPE can communicate directly with connected peers and information propagates
as connections are available. Each CPE has specific capabilities (sensing,
motion, algorithms, computing power, \ldots). CPEs are often quite simple and
are likely to be quite resource limited. A CPE needs to be situation aware, and
should function somewhat autonomously, making decisions on actions based on
local information. It should function safely even in absence of communication.
Of course a system is likely to function better with good communication and
corresponding improved knowledge of non-local state. The objective is to achieve
adaptive, robust functionality using diversity, redundancy and probabilistic
behavior. With many small entities no specific individual is critical. Entities
can come and go without disrupting the system, as long as the needed
functionality is sufficiently represented. In fact, this allows defective, worn
out, or out-of-date entities to be replaced by fresh, improved versions.

Achieving the vision of FCPS means developing new ways of thinking about how such
systems are designed and built. How are the individual entities specified so that we
can understand and predict their interactions and combined behavior? How do we design
entities that have the robustness, situation awareness needed to carryout useful
functions? What are meaningful specifications of system goals, given the distributed
nature, lack of global state, and inherent uncertainty? Rather than thinking of
correctness, we need to consider measures of satisfaction or goodness, how they can be
achieved and how they combine.

\begin{figure}[ht]
\centering
\includegraphics[width=0.67\linewidth]{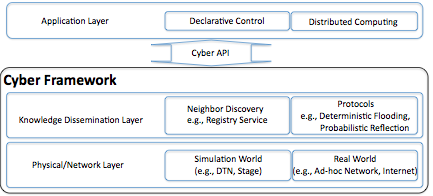}
\caption{\label{cyberfwk}
Architecture of an NCPS framework}
\end{figure}  

As a step towards the FCPS vision we have developed a framework for simulating,
emulating, and deploying networked cyber-physical systems (NCPS)
\cite{stehr-kim-talcott-10uic} (available at \url{http://ncps.csl.sri.com}). A
key feature of our NCPS framework is communication by opportunistic knowledge
exchange rather than traditional point-to-point message passing or multicast.
Each CPE can post and receive knowledge items. Knowledge can come from local
sensor readings, peer entities, or user input. Knowledge can include not only
state information (facts), but also goals that drive actions. Some goals
correspond to commands that effect the physical state or external environment,
such as turning dials, moving, or lifting. The combined knowledge of a NCPS
constitutes a distributed knowledge base. Knowledge can be exchanged between
CPEs when ever communication is possible. Under ideal conditions, all CPEs in a
system will have the same knowledge, however in general each CPE has a local
partial view consistent with a virtual global snapshot. A knowledge item can be
given a time to live, after which it is purged from local knowledge bases and
no longer propagated. This avoids cluttering the network with stale
information. There is also a replacement partial order on knowledge, allowing
one knowledge item to replace another in the knowledge base. For example, a
later measurement of position or temperature could replace an earlier one so
that the knowledge base represents the current observable state. Or, a more
important goal could replace one that is less important to focus resource use.
CPEs are free to internalize knowledge, for example to keep a history, draw
inferences and take action based on what they know locally.

An important consequence of knowledge based communication is that individual
identity no longer matters, and the number of communicating entities need not
be fixed or bounded (other than by physics). A key property for the knowledge
partial ordering is eventual consistency. In the case of a program that allows
some non-deterministic choices, but where it is important that members of a
group all make the same choice, the knowledge mechanism needs to provide a
unique conflict resolution that can be computed locally so that eventually the
local knowledge base is sufficient to detect and resolve the conflict.

Figure \ref{cyberfwk} shows the architecture of a node in our NCPS framework.
It provides a cyber API connecting the cyber part to the knowledge manager and
the physical part, which can be simulated, emulated or real. A system is a
collection of nodes that share the language in which the knowledge is
expressed. Multiple applications may run concurrently on one node, providing
flexibility and modularity. They can communicate via an internal event system,
as well as via the knowledge base. The knowledge manager is in charge of
maintaining the local knowledge base, receiving knowledge from apps and local
sensors, executing commands, notifying apps of new knowledge, and exchanging
knowledge items with other nodes. The knowledge manager implements the time to
live and replacement ordering mechanisms. Two policies for what knowledge to
exchange when are provided by the framework, and system designers are free to
add new policies. A challenging problem is to develop models of systems, their
goals and environment to guide the choice of policy that provides sufficient
propagation of knowledge without misuse of resources.


We are exploring several formalisms for specifying/programming the behavior of
CPEs including a distributed logic \cite{kim-stehr-talcott-13scp}, a
distributed work flow, and stochastic Petri Nets. Common themes are notions of
goal and degree of satisfaction, and behavior that aims to improve the degree
of satisfaction, using a utility function or similar mechanism. In these
formalisms each CPE has the same program, but may have different behavior and
actions depending on its local state and capabilities. This simplifies
programming, by managing a group of CPEs as a flexible unit. The framework
repository includes case studies exploring different aspects of design and
controlling CPEs, including various distributed mutex-like examples, sensor
robots, a distributed surveillance system, a drone system, and examples of
distributed optimization algorithms.

As an example, consider a self-organizing network of mobile robots deployed
in a building, e.g., for situational awareness during an emergency. The
robots use a common theory in the distributed logic
that specifies a language (constants,
functions, and predicates) and local inference rules based on Horn clause
logic. A robot's local knowledge (state) consists of a set of facts and a
set of goals. Facts are formulas derived by logical inference or by
\emph{observation} of the environment. Goals are formulas expressing what
the system should achieve and drive the inference process. Goals can arrive
from the environment at any time. They can also be generated as subgoals
during local inference. The communication infrastructure is such
that robots can exchange knowledge (i.e., facts and goals)
opportunistically if they reside in the same or adjacent rooms. The primary
goal is delivery of images to a specific node. The horn clause rules
describe how images are derived from
snapshots taken in an area where noise or motion is detected. Some robots
have camera devices and can move to a target area and take pictures. The
raw image may be directly sent to other nodes, or it can be preprocessed,
and then communicated to other nodes.
Primitive goals such as $TakeSnapshot(\ldots,area,image)$ correspond to 
commands that result in facts such as $Snapshot(T,area,image)$.

The knowledge partial ordering includes removal of stale facts such as
$$O1:  Position(t_P,r,\dots) \prec Position(t_P',r,\dots) ~\hbox{if}~ t_P < t_P'.$$
where $t_P,t_P'$ are time stamps, and $r$ is a robot identifier.
A new interest goal overrides ongoing tasks.  Pending subtasks are
remove using the replacement relation
$$O2: X(t_I,\ldots) \prec Interest(t'_I,\ldots)  ~\hbox{if}~ t_I < t'_I.$$ 
where $t_I, t_I'$ are time stamps also used as ``session'' identifiers
and $X$ stands for any of the subgoals generated for the earlier session.

In \cite{kim-stehr-talcott-13scp} we show that the distributed logic 
inference/execution system satisfies  \emph{Monotonicity}, \emph{Soundness} and
\emph{Completeness} properties, as well as defining conditions under which
\emph{Termination} and \emph{Confluence} hold. These are analogs of properties
of traditional inference and computation systems and are important for ensuring
desired properties of specific cyber-physical systems.

%% file: is-v-fcps.tex
\section{Immune system versus fractionated cyber-physical systems}\label{is-v-fcps}

We first consider the two systems from a requirements point of view. Then,
from another perspective we compare and contrast features and principles of
the immune system (IS) and systems based on our framework for networked
cyber-physical systems (NCPS).

\paragraph{Is the immune system a fractionated CPS?}

In some sense the immune system is the ultimate fractionated cyber-physical
system. The CPEs are cells that seamlessly combine cyber, for example
programmed decision making, with physical, for example motion, replication,
and sensing/affecting the environment via receptors and secretion. Proteins
play key roles in all aspects. There are billions of cells that sense and
affect their environment. They continually disappear (die and are degraded
and recycled---a form of garbage collection) and are replenished providing
the system with a continual supply of fresh, functioning cells. Cells
function with limited resources and act on local information. Random
specialization of T and B cells provides huge diversity.

\paragraph{Does the NCPS framework provide/support immune system features?}

As discussed in section \ref{design} key features of the immune system are: the
two-level architecture, self-evaluation and monitoring, pattern matching
foundation for interaction and for safety/security mechanisms. The two-level
architectures is very much in the spirit of FCPS, in that a system may include
CPEs of different capabilities, response time, effectiveness such as elements
of the innate and adaptive immune system. Self monitoring can be provided by
adding a monitor application to the collection of applications running on a
node, with out changing other applications. The monitor can use the shared
knowledge and events to evaluate system state. The effectiveness of the monitor
depends on the design of the knowledge system, the ability to sense and
intervene provided by the physical system, and interruptibility of other
applications. Event based execution facilitates the latter. These both are
interesting design patterns that could be formalized in the NCPS framework.

Communication, safety, and security in the immune system is founded on pattern
matching. Some patterns are essentially unforgeable, similar to physical finger
prints or signatures. The use of largely unforgeable patterns for
authentication, access control, and provenance in another intriguing topic for
future research. For a given class of applications, can we find a pattern
algebra/system that provides the right security primitives? How can they
realized, perhaps by a combination of cyber and physical elements? What are
potential attack models and how do we prevent attacks or mitigate the result?
Can we make attacks too expensive to be attractive? This could be a new
approach to securing pervasive and adaptive NCPS.

\paragraph{IS principles for NCPS.}

We listed a number of principles that the immune system 
has evolved to obey.  Are these realized/realizable in FCPS?

\emph{Diversity} happens at muliple levels. Random choice of antibody/receptor
construction in the immune system provides huge diversity in the invader
patterns that can be matched. Random choice of parameters or branches in a
workflow provides diversity in in the NCPS framework. Interestingly a cell
makes the choice in advance and each choice is represented concurrently in the
system, while a CPE makes the choice dynamically, requiring fewer CPEs to cover
all bases. The immune system provides multiple mechanisms for activating
response and for clearing invaders and damages cells. \emph{Multiple levels of
specificity} is another form of diverisity. These are reflected in the FCPS
philosophy of multiple capabilities, and redundancies to enable different
solutions to a goal. The immune system can serve as a first model for designing
such systems.

\emph{Clonal selection} is the principle that selects only the T and B
cells with receptors recognizing an invader for replication. This could be
realized as a form of program specialization and replication (across
multiple cores) in response to an urgent need---in response to an attack or
other emergency, or an unusual computation load.

\emph{Safeguards} include multi-ary control and thresholds for action. The
use of utility functions and the notion of degree of satisfaction used to
program CPEs can be compared to the fact that some cells, such as T cells,
require the signal intensity reach a certain threshold before they become
active. \emph{Fault tolerance} is achieve by monitoring, fitness tests, and
designs that ensure that one fault not enough to succeed, for example a
rogue CTC cell recognizing self is unlikely to simultaneously see the
additional patterns required for it to take action. Monitoring and
self-checking were discussed above. Cross checks and multi-key
access/actions are interesting patterns to be included in an FCPS library.

\emph{Mobility} is controlled using potential gradients and patterns as
signposts. The use of objective/utility functions to improve goal
satisfaction by a CPE can be compared to the use of biomolecules, such as
chemokines, that generate a chemical potential gradient to attract certain
types of cell. Current experimental setups for robot testing use external
cameras, or RFID tags in the floors and on walls for robots to locate
themselves. GPS can be used for coarse grained location. The use of
patterns to control navigation and identify target locations is an
intriguing alternative, perhaps more natural and closer to how humans
navigate.

\paragraph{Other issues.}
In both cases there is a common language across CPEs (DNA/proteins or
knowledge representation) and one program (DNA vs logical theory or
workflow) from which multiple roles emerge according to local state and
capabilities. Programmed cell death can be compared to time to live for a
knowledge item as a means of removing stale information. Both systems have
multiple modes of propagating information: peer-to-peer (cell-cell
binding), broadcast (secretion), and courier (mobility).

NCPS uses a partial order on knowledge as a mechanism for controlled forgetting
or over-riding and knowledge is uniformly shared across all CPEs. In the immune
system, received signals are processed and new signals generated. Thus
``knowledge'' is consumed with no need for replacement.
There is no attempt to form a global picture.  Each entity needs and
only makes use of locally available information.   An interesting
question is whether/how specific classes of NCPS can be designed
to achieve such locality in information needs.

%% file: conclusion.tex
\section{Conclusion}\label{conclusion}

We have presented highlights of the human immune system from a computer science,
distributed systems point of view, summarized the key features and design
principles and compared this to the notion of fractionated cyber-physical
systems (FCPS). As discussed in section \ref{is-v-fcps} there are many
similarities. At the qualitative level we see two key differences: the use of
pattern matching and the safety and security mechanisms of the immune system.

The immune systems is viewed and modeled primarily as a defense system, although
immune system components appear to have roles in the functioning of other
biological systems. Although FCPSs are not necessarily defense systems, the
distributed and open nature requires that defense mechanisms be built into the
supporting infrastructure to be used as appropriate. Thus we should think about
the immune system features and principles in term of more general notions of
threat or fault and goals or objectives and how/whether the security/safety
mechanisms work at an abstract level.

One can think of pattern matching as analogous to cryptographic mechanisms
such as decryption and signature checking.  There is also a flavor of
role or attribute based access control, since cells playing different roles
express different receptors (attributes) and a cell can only receive a
signal (read a message) if it expresses the corresponding receptor.
Perhaps some form of attribute based encryption could be used 
to support flexible security policies in the NCPS framework.

An interesting topic for investigation is whether there are cyber-physical
analogues to the ability of cells to express unique patterns in a relatively
unforgeable manner that could be used as the basis for building safety/security
mechanisms for FCPSs.  Finger prints and DNA identify individuals, but we
really want to determine trustworthiness of classes of CPEs and to reliably
identify threats or obstacles.

We leave the reader with a grand challenge: 
\emph{What is the mathematics of the immune system control?} This is a
fascinating phenomenon. The answer could lead to new ways to design and manage
FCPSs. Such a mathematical model should explain how important features of the
dynamic behavior can be derived from locally determined behavior. It must
account not only for individual response to locally available information but
also how information propagates. It must also account for the openness of
such systems and the uncertainty. Factors include the safety and security
mechanisms as well as well as the signaling patterns, signaling and
reproduction/reinforcement rates, and the various positive and negative feedback
loops sketched at the end of section \ref{is}. There are many trade offs,
effects to balance and often a need for rapid response and adaptation.
\begin{itemize} 
\item What is the trade space for designs? 
\item How can regions of undesirable behavior be predicted? 
\item How can regions of desirable behavior be ensured? 
\end{itemize}